\documentclass[prb,twocolumn,superscriptaddress]{revtex4}

\usepackage{graphicx,epsfig}
\def\figwidth{9cm}

\begin{document}

\title{Spinless fermions ladders at half filling}

\author{P. Donohue}
\email{donohue@lps.u-psud.fr} \affiliation{Laboratoire de
Physique des Solides, CNRS-UMR 85002, Universit\'e Paris--Sud,
B\^at. 510, 91405 Orsay, France}

\author{M. Tsuchiizu}
\email{tsuchiiz@edu2.phys.nagoya-u.ac.jp} \affiliation{Department
of Physics, Nagoya University,
             Nagoya 464-8602, Japan}

\author{T. Giamarchi}
\email{giam@lps.u-psud.fr} \affiliation{Laboratoire de Physique
des Solides, CNRS-UMR 85002, Universit\'e Paris--Sud, B\^at. 510,
91405 Orsay, France}

\author{Y. Suzumura}
\email{e43428a@nucc.cc.nagoya-u.ac.jp} \affiliation{Department of
Physics, Nagoya University,
             Nagoya 464-8602, Japan}
\affiliation{CREST, Japan Science and Technology Corporation
(JST), Japan }

\date{\today}

\begin{abstract}
We study a half filled ladder of spinless fermions. We show that
contrarily to a single chain, the ladder becomes a Mott insulator
for arbitrarily small repulsive interactions. We obtain the full
phase diagram and physical quantities such as the charge gap. We
show that there is only a single insulating phase for repulsive
interactions, regardless of the strength of the interchain hopping
and single chain Mott gap. There is thus no
confinement-deconfinement transition in this system but a simple
crossover. We show that upon doping the system becomes a Luttinger
liquid with a universal parameter $K=1/2$ different from the one
of the single chain ($K=1/4$).
\end{abstract}

%\maketitle must follow title, authors, abstract and \pacs
\maketitle

\section{Introduction}

One dimensional systems are one of the few known example of
non-fermi liquid
behavior\cite{schulz_houches_revue,voit_bosonization_revue,gogolin_1dbook}.
It is thus of utmost theoretical importance to understand how one
can go from a one dimensional situation to a more conventional
high (typically three) dimensional one by coupling one
dimensional systems, allowing particles to jump from chain to
chain. This question is far from being elucidated despite several
theoretical attempts\cite{bourbonnais_couplage,wen_coupled_chains,yakovenko_manychains,%
clarke_coherence_coupled,schulz_moriond,georges_organics_dinfiplusone}.
For commensurate one dimensional system another phenomenon
appears: such systems are Mott insulators. This leads to a direct
competition between interactions and hopping. The insulating
behavior of the one dimensional system tends to kill the
interchain hopping and thus to confine the electrons on individual
chains. Conversely, a large interchain hopping destroys the
one-dimensional character and thus weakens the Mott transition
considerably, turning the system into a metal. This competition
between the Mott transition and interchain hopping has in addition
to its theoretical importance, implications for organic compounds
\cite{giamarchi_mott_ref,vescoli_confinement_science,moser_conductivite_1d,henderson_transverse_optics_organics}
that are three dimensional stacks of quarter filled
chains\cite{jerome_revue_1d}.

Unfortunately studying an infinite number of coupled chains is
extremely difficult, so to understand such phenomenon it is
interesting to investigate simpler systems with a finite number of
coupled chains. Such systems are the so-called
ladders\cite{dagotto_ladder_review}. They present the advantage to
allow a careful study of the effects of hopping by being tractable
by powerful
analytical \cite{fabrizio_2ch_rg,kveschenko_spingap,finkelstein_2ch,%
schulz_2chains,balents_2ch,nagaosa_2ch,nersesyan_2ch,yoshioka_coupledchains_interaction}
and numerical techniques
\cite{dagotto_lanczos_2ch,noack_dmrg_2ch,poilblanc_2ch_mc,poilblanc_2ch,%
tsunegutsu_2ch}. For commensurate ladders with spin, the relevance
of interchain hopping was studied by renormalization group
techniques
\cite{suzumura_confinement_ladder,tsuchiizu_confinement_ladder,tsuchiizu_confinement_ladder_long}.
Depending on the ratio between the single chain Mott gap and the
interchain hopping (suitably renormalized by the interactions) a
very different flow of the single particle hopping was observed,
reminiscent of the confinement-deconfinement transition expected
for the infinite number of chains, even if in the ladder there is
no real transition but a simple crossover
\cite{lehur_ladder_crossover}. In addition, interchain hopping was
shown to drastically modify the critical properties of the Mott
transition
\cite{schulz_mitwochain,lin_so8,konik_exact_commensurate_ladder}
compared to the one of a single chain. Despite these studies on
commensurate ladders a detailed description of the phase diagram
and of the nature of the Mott transition is still lacking.

In the present paper we investigate these issues on a ladder of
spinless fermions. Spinless fermions exhibit extremely interesting
behavior since a single chain needs a finite repulsive interaction
before turning into a Mott insulator, contrarily to the spinful
chain for which any repulsive interaction freezes the charge
leaving only the spin degrees of freedom. One could thus naively
think to be able to go from an insulating phase, dominated by the
single chain gap, to a metallic phase even for repulsive
interactions. In fact, quite interestingly, for the spinless
ladder the Mott transition is pushed in the vicinity of the
non-interacting point, invalidating this naive picture. Quite
fortunately the fact that the Mott transition is now in the
vicinity of the non interacting point allows to study it using
standard renormalization group technique, and extract the complete
properties of the transition.

The plan of the paper is as follows. In section~\ref{sec:model} we
introduce the model for the two leg spinless ladder. In
section~\ref{sec:rg} we study this model using the renormalization
group technique. We show that the Mott transition occurs now for
arbitrary repulsive interactions and compute the various physical
parameters (charge gap, Luttinger liquid parameters) both
analytically and by a numerical integration of the RG equations.
We analyse the phase diagram in section~\ref{sec:phasediag}. We
show that in the ladder the confinement-deconfinement is in fact a
crossover. We also investigate the properties of the slightly
doped ladder and point out the differences that exist compared to
a doped single chain. Conclusions can be found in
section~\ref{sec:conclusion}. Finally some technical details can
be found in the appendix.

\section{Model}\label{sec:model}

We start from spinless electrons on a two leg ladder, described by
the Hamiltonian
\begin{eqnarray}
H  &=&  -t \sum_{i,\alpha} (c^{\dagger}_{i,\alpha} c_{i+1,\alpha} +
\text{h.c.}) + V \sum_{i,\alpha} n_{i,\alpha} n_{i+1,\alpha}
\nonumber \\
& & - t_\perp \sum_{i} (c^{\dagger}_{i,1} c_{i,2} + \text{h.c.})
\label{teve}
\end{eqnarray}
where $\alpha=1,2$ is the chain index. $t$ and $t_\perp$ are
respectively the intra and interchain hopping, and $V$ the
repulsion between nearest neighbors particles.

To analyze the long distance properties of this model it is
convenient to use the boson representation of fermions operators
\cite{schulz_houches_revue,voit_bosonization_revue,gogolin_1dbook},
valid in one dimension. Two basis are possible: (i) one can start
in the original chain basis and bosonize each chain; (ii) one can
use the bonding and antibonding basis. Each basis has advantages
and drawbacks and we will need both to tackle the Mott transition
in the ladder, so we give both boson representations below.

\subsection{Chain basis}

We refer the reader to the literature for the boson mapping and
recall here only the main steps to fix the notations. Taking a
linearized energy dispersion at the Fermi level, we use the
following expressions for right and left moving fermions
\cite{schulz_houches_revue}.
\begin{equation}
\Psi_{R,L} = \frac{\eta_{R,L}}{\sqrt{2 \pi a}}
e^{-i(\pm\phi-\theta)}e^{\pm i k_F x}
\end{equation}
$\phi$ and $\Pi$ are canonically conjugate operators and
$\Pi=\frac{1}{\pi}\partial_x{\theta}$. $\nabla \phi$ and $\nabla
\theta$ give respectively the long wavelength fluctuations of the
density and current. We note $\eta_{R,L}$ Klein factors which one
must introduce to reproduce the anticommutation properties of
several fermion species\cite{schulz_moriond}.
$a$ is a short distance cutoff of the order of the lattice spacing.
With these operators the single chain Hamiltonian takes the form:
\begin{eqnarray} \label{limcontinue}
H &=& H_0 + H_{int} + H_u \\
H_0 &=& \frac{v_F}{2\pi} \int dx \left[(\pi \Pi)^{2} +
(\partial_{x} \phi)^{2}\right] \nonumber \\
H_{int} &=& g \int dx (\partial_{x} \phi)^{2} \nonumber \\
H_u&=&-2g_{u}\int\frac{dx}{(2\pi a)^2} \cos(4\phi) \nonumber
\end{eqnarray}
where $v_F$ and $g$ are respectively the bare Fermi velocity and
the interaction. They are given by
\begin{eqnarray}
v_F &=& 2ta\sin(k_F a) \\ g &=& \frac{(1-\cos(2k_F a)) aV}{\pi^{2}}
\end{eqnarray}
The umklapp part, $H_u$, only appears in this form at half-filling
\cite{giamarchi_mott_ref}(i.e. when $4k_Fa =2\pi$)  and is
responsible for the Mott transition of a single chain. The
interaction $g$ can be absorbed in the quadratic part to give the
Luttinger Hamiltonian
\begin{equation}
H = \frac{1}{2\pi} \int dx \left[u K(\pi \Pi)^{2} +
\frac{u}{K}(\partial_{x} \phi)^{2}\right]
\end{equation}
The parameters of the Hamiltonian are the renormalized fermi
velocity $u$, the Luttinger $K$ parameter and the non-universal
umklapp coupling constant. For small $V$ these parameters can be
perturbatively computed:
\begin{eqnarray}
u K &=& v_F \nonumber \\ \frac{u}{K} & = & v_F +
aV\frac2\pi(1-\cos(2k_F a)) \\ g_u &=& aV   \nonumber
\end{eqnarray}
However the description (\ref{limcontinue}) is much more general
and is valid even at large coupling provided the proper
renormalized coupling constant are used. In fact in the following
we will not assume such relations for the coupling constants and
take $g_u$ as a free parameter (the parameters may be tuned at
will with for example a second nearest-neighbor interaction).

Using the bosonized expression (\ref{limcontinue}) for the single
chain, we can write the two uncoupled chains in (\ref{teve}) as
\begin{eqnarray}\label{eq:hint}
H_{int} &=&  g\int dx ((\partial_{x} \phi_s)^{2}+(\partial_{x}
\phi_a)^{2})  \\ H_u &=& -4g_{u}\int \frac{dx}{(2\pi a)^{2}} \cos
( \sqrt{8}\phi_a )\cos ( \sqrt{8}\phi_s ) \nonumber
\end{eqnarray}
where we define
\begin{eqnarray}
\phi_s &=& \frac{1}{\sqrt{2}}(\phi_1+\phi_2) \\ \phi_a &=&
\frac{1}{\sqrt{2}}(\phi_1-\phi_2)
\end{eqnarray}

The interchain hopping in (\ref{teve}) reads in this basis
\begin{equation}
\label{tperpcont} -t_\perp \frac{2}{ \pi a } \cos{\sqrt{2}\phi_a }
\cos{\sqrt{2} \theta_a }
\end{equation}

This basis has the advantage to treat very simply the umklapp term,
but has the drawback not to reproduce easily the band picture of free fermions.

\subsection{Two band basis}

Another basis is the bonding-antibonding band basis. We first
diagonalize the kinetic energy in (\ref{teve}) with
\begin{eqnarray}
c_{i,0} = \frac1{\sqrt{2}} (c_{i,1}+c_{i,2}) \\ c_{i,\pi} =
\frac1{\sqrt{2}} (c_{i,1}-c_{i,2})
\end{eqnarray}
and the corresponding boson fields. In this basis the interchain
hopping is diagonal:
\begin{equation} \label{tperpband}
-t_{\perp} \sum_{i}(c^{\dagger}_{i,0}c_{i,0}-
c^{\dagger}_{i,\pi}c_{i,\pi})
\end{equation}
however the interaction term is less simple to formulate. Rather
than to use the bonding and antibonding boson fields it is again
convenient to introduce the symmetric and antisymmetric
combination that we denote now $\phi_\rho =(\phi_0 +
\phi_\pi)/\sqrt{2}$ and $\phi_\sigma =(\phi_0 -
\phi_\pi)/\sqrt{2}$ to distinguish them from the ones in the chain
basis. This leads to the simple bosonized expression:
\begin{equation}\label{tperpbandcont}
-\frac{t_{\perp} \sqrt{2}}\pi \int dx \partial_{x} \phi_\sigma
\end{equation}
The change of basis for the interaction term can be most easily
performed using the transformation formulas for  the total charge
current .
\begin{eqnarray} \label{changebasis}
-1/\pi \partial_{x} \phi_s &=& -1/\pi \partial_{x} \phi_\rho \\
-1/\pi \partial_{x} \phi_a &=& \frac{\sqrt{2}}{ \pi a }
\cos{\sqrt{2} \phi_\sigma } \cos {\sqrt{2} \theta_\sigma }
\nonumber
\end{eqnarray}

Since the interaction terms are quadratic in currents we also need
an operator product expansion to extract the most relevant
operators.
\begin{equation}\label{ope}
\cos{n \phi}^2 = \frac12 (1 - \frac{n^2 a^2}{2} (\partial_{x}
\phi)^2 + \cos{2n \phi})
\end{equation}
The expression for the anti-symmetric current is however
misleading since it does not include Klein factors required to
identify bosonic exponents to anticommuting
fermions\cite{schulz_moriond}. To obtain the exponential terms it
is necessary to read the bosonized expression on the transformed
four fermion operators\cite{giamarchi_spin_flop}. With these boson
operators the interaction term takes the following
form\cite{nersesyan_2ch}:
\begin{eqnarray}
H_{int} &=& \frac{2 V a}{\pi^{2}}\int dx (\partial_{x}
\phi_{\rho})^{2}  \\
 & & - \frac{V a}{ \pi^2} \int dx (\pi^2 \Pi_\sigma^2 + (\partial \phi_{\sigma})^2) \nonumber \\
 & &+ \frac{V a}{\pi^2} \int \frac{dx}{a^2} (\cos{\sqrt{8}\theta_{\sigma}}
 -\cos{\sqrt{8}\phi_{\sigma}} \nonumber \\
 & & -\cos{\sqrt{8}\phi_{\sigma}} \cos{\sqrt{8}\theta_{\sigma}} )\nonumber
\end{eqnarray}
and the umklapp term is
\begin{equation}
H_u =  \frac{-g_{u}}{2\pi^{2}a^2} \int dx
\cos(\sqrt{8}\phi_{\rho})(\cos(\sqrt{8}\phi_{\sigma})
 +\cos( \sqrt{8} \theta_{\sigma}))\nonumber
\end{equation}
We now define the coupling constant $g_\sigma$ associated with the
operator $\cos(\sqrt8\theta_\sigma)$.
\begin{equation}
\delta H= 2g_\sigma \int \frac{dx}{(2\pi a)^2}\cos{\sqrt8 \theta_\sigma}
\end{equation}

\section{Mott transition} \label{sec:rg}

Let us now analyse the Mott transition. The single chain needs a
finite strength interaction in order to become an insulator.
Indeed, as can be seen from (\ref{limcontinue}), for a single
chain the umklapp term has a dimension $2-4K$. It thus opens a gap
in the charge sector and leads to a Mott insulating phase for
$K<1/2$ and a metallic (Luttinger liquid) phase for $K>1/2$. For
the t-V model this corresponds to $V=2t$. In the ladder the
presence of interchain hopping dramatically modifies this. As we
will show, the ladder is an insulator for infinitesimal repulsive
interactions. This can be easily seen by looking in the chain
basis at the operators generated by $t_\perp$. Obviously $t_\perp$
will generate terms such as $\cos(\sqrt{8}\phi_a)$ and
$\cos(\sqrt{8}\theta_a)$. Note that these terms only contain the
antisymmetric field. However when combined with the umklapp
(\ref{eq:hint}) the $\cos(\sqrt{8}\phi_a)$ will generate a
$\cos(\sqrt{8}\phi_s)$. This term has the dimension $2-2K_s$, and
in contrast with the single chain umklapp, is relevant for $K_s
<1$ leading to a gap in the symmetric sector and hence to a Mott
insulating phase. The perpendicular hopping thus reinforces the
insulating character of the system.

To go beyond this simple argument let us now investigate the full
RG flow using a two scale analysis of the relevant operators in
the Hamiltonian.

\subsection{Mott insulator for $1/2<K<1$} \label{sec:deconfined}

Let us first focus for values of $K$ for which the single chain
would be metallic. In that case single chain umklapp operator
(\ref{limcontinue}) is irrelevant. The gap in this regime
thus results from the competition of the initial decrease of the
umklapp constant in the flow and its subsequent growth after
$t_\perp$ has blocked the transverse fluctuations.

To investigate the gap let us write the full RG equations, which
include the umklapp term. Near the non-interacting point, if
$t_\perp$ is small the flow is given by
\begin{eqnarray}\label{eq:flow2chain}
\frac{dg_u}{dl} &=& (2-2K_a-2K_s)g_u \nonumber
\\ \frac{dg_s}{dl} &=& (2-2K_s)g_s+\frac{1}{\pi} g_u g_a \nonumber \\
\frac{dg_a}{dl}&=&(2-2K_a)g_a+\pi (K^{-1}-K)
(t_\perp a)^2+\frac{1}{\pi} g_u g_s \nonumber
\\ \frac{dg_f}{dl} &=&(2-2/K_a)g_f+\pi (K^{-1}-K)(t_\perp a)^2 \nonumber \\
\frac{dK_a}{dl}&=&\frac{1}{2\pi^2}(g_f^2-g_a^2-2g_u^2) \nonumber
\\ \frac{dK_s}{dl}&=&\frac{-1}{2\pi^2}(g_s^2+ 2g_u^2) \nonumber \\
\frac{dt_\perp}{dl}&=&(2-\frac{1}{2}(K_a+1/K_a))t_\perp
\end{eqnarray}
where we have set $g/u \to g$ ($u=1$) to keep the notations simple.
For $g_u=0$ these
equations reduce to the ones obtained in
Ref.~\onlinecite{nersesyan_2ch}. We have introduced couplings
which are generated during the flow, though they are not present
in the bare hamiltonian:
\begin{eqnarray}
\delta H&=&2g_a\int \frac{dx}{(2\pi a)^2}\cos{\sqrt8 \phi_a}
-2g_f\int \frac{dx}{(2\pi a)^2}\cos{\sqrt8 \theta_a}\nonumber \\ &
& +2g_s\int \frac{dx}{(2\pi a)^2}\cos{\sqrt8 \phi_s}
\end{eqnarray}
These new couplings are $g_a$  which is  a density-density
interaction between the two chains and $g_s$ which is the umklapp
part of this interaction. The coupling $g_f$ transfers a pair of
fermion from one chain to the other and corresponds to a Josephson
coupling between the two chains.

For a finite value of the renormalised $t_\perp$ that  depends on
the initial fermi velocity and interactions, one can neglect the
interaction terms that couple the two bands in a non-resonant way.
Hence we can use the above flow equations up to a lengthscale
$l_1$ where $t_\perp$ reaches this finite value, $t_\perp \sim
O(1)$. Above the scale $l_1$, it is more convenient to switch to
the two band basis in order to study the flow. By definition of
$l_1$ we discard any term that contains $\cos{\sqrt8
\phi_\sigma}$, since such terms transfer momentum among the bands.
Above $l_1$ the flow becomes
\begin{eqnarray}
\frac{dK_\sigma}{dl}= \frac{1}{2\pi^2}(g_\sigma^2+\frac12 g_u^2) \nonumber \\
\frac{dg_\sigma}{dl}=(2-2K^{-1}_\sigma)g_\sigma
\label{eq:flowlarge}
\\ \frac{dg_u}{dl}=(2-2K_\rho-2K_\sigma^{-1})g_u \nonumber
\end{eqnarray}
In the $\sigma-$sector the remaining $\cos{\sqrt{8}\theta_\sigma}$
term  opens a gap. We note $l_\sigma$ the scale when the coupling
term $\cos{\sqrt{8}\theta_\sigma}$ has flowed to a value of order
one, and the gap amplitude is evaluated as $\Delta_\sigma=t
e^{-l_\sigma}$.
Above the scale $l_\sigma$ the umklapp operator becomes relevant
since it is reduced to a simple $\cos{\sqrt{8}\phi_\rho}$.
\begin{equation} \label{eq:flowult}
\frac{dg_u}{dl} = (2-2K_\rho)g_u
\end{equation}
These equations describe completely the Mott transition in the ladder system.
An analytical solution can be given both in the limit of very small interchain
hopping and for large interachain hopping but in the limit of very small interactions.
A numerical integration
of the equations allows to obtain the gap for arbitrary initial parameters.

\subsubsection{$t_\perp\to 0$}

Since the
umklapp is irrelevant for a single chain, one can replace the
flow, when $t_\perp$  is the smallest scale in the problem, by
\begin{eqnarray}
\frac{dg_u}{dl} = (2-4K^*)g_u \nonumber \\
\frac{dt_\perp}{dl}=(2-\frac{1}{2}(K^*+1/K^*))t_\perp
\end{eqnarray}
where $K^*$ is the renormalized Luttinger parameter for a single
chain. For the $t-V$ model $K^*$ can be obtained exactly
\cite{haldane_xxzchain,luther_chaine_xxz}. When $t_\perp \sim O(1)$ we switch to the
second set of flow equations (\ref{eq:flowlarge}). We do not know
the precise renormalization of the umklapp constant between $l_1$
and $l_\sigma$. But since $l_\sigma - l_1$ does not depend on
$t_\perp$, this gives a simple multiplicative constant. Above $l_\sigma$
we use (\ref{eq:flowult}). Integration of the RG equations yields:
\begin{eqnarray}
g_u(l_1) = g_u(0) e^{(2-4K^*)l_1}\nonumber\\
 g_u(l_\sigma)=g_u(l_1)e^{\beta (l_\sigma-l_1)}\nonumber\\
g_u(l_\rho)=g_u(l_\sigma)e^{(2-2K^*)(l_\rho-l_\sigma)}
\end{eqnarray}
We have used $\beta$ as a constant that takes into account the
variation of $K_\sigma$ between $l_1$ and $l_\sigma$ as it flows
to zero. This does not change the $t_\perp$ dependence. Collecting
the intermediate results, we can give an asymptotic dependence of
the total charge gap on $t_\perp$:
\begin{equation}
\Delta_\rho \propto
t_\perp^{\frac{K^*}{1-K^*}\frac{1}{2-\frac{1}{2}(K^*+1/K^*)}}
\end{equation}
While deriving this result we have neglected terms generated to
second order by $t_\perp$, such as $g_a$,$g_f$ and $g_s$. In the appendix \ref{ap:alternative}
we show
that including those terms do not affect the dependence of
$\Delta_\rho$ on $t_\perp$ .

\subsubsection{$V\to 0$, large $t_\perp$}

Another interesting limit is when $t_\perp$ is comparable to $t$.
Note that we always remain in the limit where $t_\perp < t$ in
order to keep four points at the Fermi level, otherwise the
problem would be the trivial one of a single filled band of
fermions. In that case the initial flow does not exist and we
start directly with (\ref{eq:flowlarge}).  The umklapp operator
has an initial dimension of $2-4K$ and thus initially decreases in
the flow. However combined with the operator $g_\sigma\cos{\sqrt8
\theta_\sigma}$ it generates from second order in the interaction
expansion a term of the form $g_\rho \cos{\sqrt8\phi_\rho}$. This
new operator is relevant, its scaling dimension being $2-2K_\rho$.
The flow equation of this operator reads:
\begin{equation}
\frac{dg_\rho}{dl}=(2-2K_\rho)g_\rho +\frac{1}{2\pi} g_u g_\sigma
\end{equation}
Using the flow equations at large $t_\perp$ (\ref{eq:flowlarge}),
we have $g_u(l)\approx Ve^{-(2-V)l}$ and $g_\sigma(l)=V+O(V^2l^2)
$ , these expressions are valid at the beginning of the flow and
we have taken into account only the leading dependence on V. This
leads to an approximate expression for $g_\rho$, using the fact
that for small interactions $2-2K_\rho \propto V$ :
\begin{equation}\label{eq:grho}
g_\rho(l)=\frac{V^2}{2}(e^{\frac{2V}{t\pi}l}-e^{-2l})
\end{equation}
The coupling $g_\rho$ starts at zero and is driven by $g_u$ which
is decreasing rapidly. We may now determine the scale $l^*$ where
$g_\rho$  becomes larger than $g_u$. Using (\ref{eq:grho}) one
gets:
\begin{equation}
\frac{V^2}{2}e^{Vl^*}=Ve^{-(2-V)l^*}
\end{equation}
which yields $l^*\propto
\ln{\frac{1}{V}}$. We note that at this scale $g_\sigma$ has
hardly changed at all, which validates the expression for
$g_\rho$. Indeed $g_\sigma$ reaches a value of order one at
$l_\sigma\propto\frac{1}{V}$ which for asymptotically small
interactions is much larger than $l^*$. The charge gap is dominated
by $g_\rho$ and we may safely
drop the original umklapp operator which gives only subdominant contributions.
The gap in the total charge
sector is given by the scale where $g_\rho$ reaches a value of
order one. This gives a gap
\begin{equation}
\ln{\Delta_\rho}\propto %\pi t
\frac{\ln{V}}{V}
\end{equation}
We note that the gap in the charge sector decreases faster than
the gap in $\sigma-$sector, where $\ln{\Delta_\sigma}\propto
V^{-1}$.

\subsubsection{Numerical solution of the equations}

\begin{figure}
\includegraphics[width=9cm]{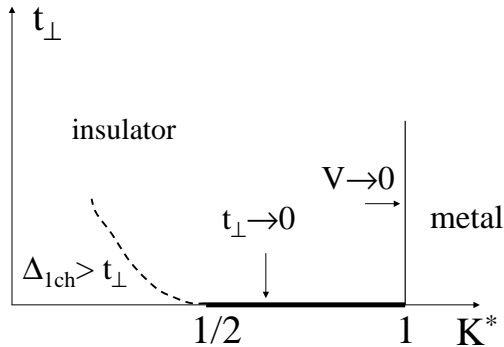}
\caption{\label{fig:phasediag} Phase diagram as a function of
interaction and interchain hopping ($t_\perp$). The ladder is an
insulator for any repulsive interaction ($K^*<1$), while the
single chain is insulating for $K^*<1/2$ . The dashes represent a
cross-over between a region where $t_\perp$ is relevant in the RG
sense and a region where it is irrelevant ($\Delta_{1ch}$ is the
gap on a single chain). For $1/2<K^*<1$ the charge gap, $\Delta_\rho$, vanishes
as a power law of $t_\perp$ for small $t_\perp$ and as
$\ln{\Delta_\rho}\propto
{\ln{V}}/{V}$ for small interaction (see text).}
\end{figure}

\begin{figure}
\centerline{\epsfig{file=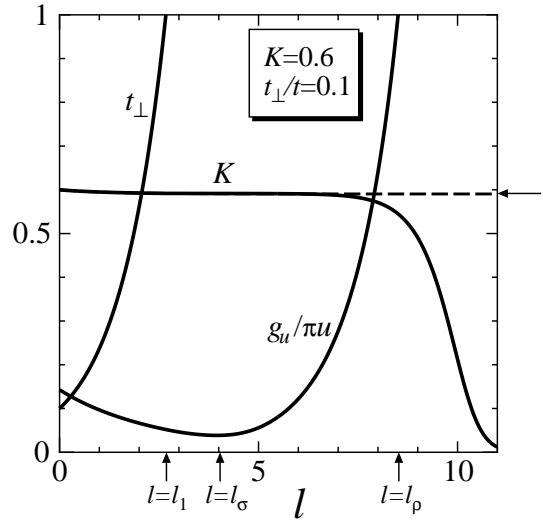,angle=0,width=\figwidth}}
\caption{\label{fig:rgflow1} Typical RG flow, for $1/2<K<1$,
showing the two regime: below $l_1$
%  (scale where $t_\perp$ is of order one)
  (scale where $t_\perp$ becomes unity)
and above where some couplings are cut by  momentum
conservation enabling the umklapp term to flow to strong coupling.
The quantities $l_\sigma$ and $l_\rho$ show the scales where
  $g_\sigma/\pi u$ and $g_\rho/\pi u$ become  unity
  at $l=l_\sigma$ and $l=l_\rho$ respectively.
The dashed curve represents the flow for $t_\perp=0$ and
  the arrow denotes $K^*(\simeq 0.59)$.
}
\end{figure}

\begin{figure}
\centerline{\epsfig{file=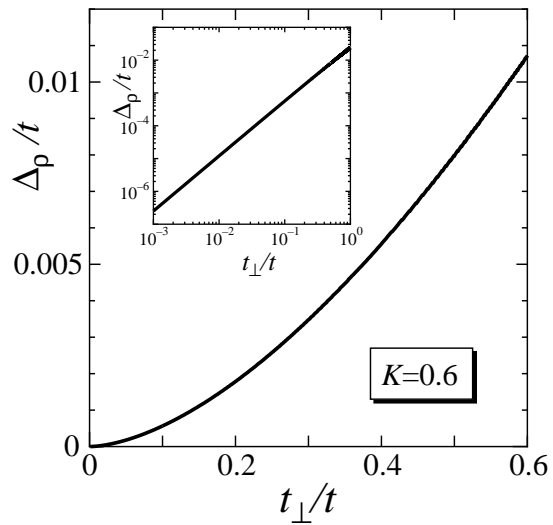,angle=0,width=\figwidth}}
\caption{\label{fig:chargegaptperp} Charge gap as a function of
$t_\perp$ for finite interactions characterized by the Luttinger
parameter $K$. The inset shows the power law behavior of the
charge gap, for small interchain hopping.}
\end{figure}

These asymptotic behaviors and the phase diagram are shown in
Figure~\ref{fig:phasediag}.
To go beyond the asymptotics either at small $t_\perp$ or at the
transition $V\to 0$, one needs to numerically integrate the flow
(\ref{eq:flow2chain},\ref{eq:flowlarge},\ref{eq:flowult}). One
example of such a flow is shown on Figure~\ref{fig:rgflow1}.
One clearly sees the initial decrease of the umklapp in the
initial phases of the flow, followed by its subsequent increase
when $g_f \sim O(1)$. The charge gap is given in
Figure~\ref{fig:chargegaptperp}.

\subsection{Gap for $K<1/2$} \label{sec:confined}

\begin{figure}
\centerline{\epsfig{file=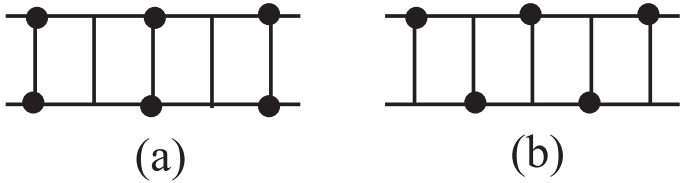,angle=0,width=\figwidth}}
\caption{\label{fig:cdwlock}Degenerate ground states for two
uncoupled chains, (a) and (b). The interchain hopping selects the
out of phase configurations (b). Such a ground state is the same
in the whole insulating phase, regardless of whether $\Delta_{\rm 1ch}> t_\perp$
or not (see text).}
\end{figure}

\begin{figure}
\centerline{\epsfig{file=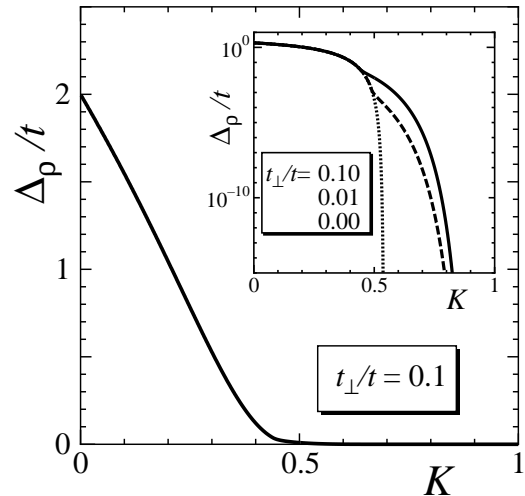,angle=0,width=\figwidth}}
\caption{\label{fig:fullgap} Charge gap for finite $t_\perp$ as a
function of the Luttinger parameter $K$, the inset shows curves
for different values of $t_\perp$. For $K<1/2$ the Mott gap is
essentially the single chain gap, whereas it is strongly $t_\perp$
dependent for $1/2<K<1$.}
\end{figure}

The picture is quite different in this case since without
interchain hopping a gap would be present on each chain. For small
$t_\perp$ the model is conveniently studied in the chain basis,
 using the flow (\ref{eq:flow2chain}).
Because the field $\phi_{1,2}$ orders, the single particle hopping
between the chains is irrelevant (for small $t_\perp$). The only
remaining relevant coupling is thus the generated interchain
density-density interaction $g_a$. This coupling opens a gap in
the antisymmetric sector and orders $\phi_a$. The physics is quite
clear: in the absence of interchain coupling each chain has a
charge density wave ground state that is double degenerate. In the
presence of $t_\perp$, particles still cannot hop from chain to
chain because of the Mott gap in each chain, but the virtual hops
tend to lock the two CDW relative to each other. This is shown on
Figure~\ref{fig:cdwlock}.
The Mott gap is thus essentially here the single chain gap.
The gap as a function of $K$ obtained by numerical integration of
the flow is shown in Figure~\ref{fig:fullgap}.

\section{Phase diagram}\label{sec:phasediag}

The results of section~\ref{sec:rg} allow to draw the phase
diagram of the commensurate ladder, as shown on
Figure~\ref{fig:phasediag}. The very existence of an insulating
phase in the ladder for $1/2<K<1$ prompts for several questions.

\subsection{Confinement vs crossover}

Apparently we have to face two very different behaviors depending
on the strength of $t_\perp$. For $K<1/2$ and small $t_\perp$,
$t_\perp$ renormalizes to zero as was shown in
section~\ref{sec:confined}, whereas for large $t_\perp$ (or for
$1/2<K<1$ where the single chain gap is absent) $t_\perp$
renormalizes to large values. One thus seems to have a
confinement-deconfinement transition, similar to the one expected
for an infinite number of chains, induced by the competition
between $t_\perp$ and the single chain Mott gap $\Delta_{\rm
1ch}$. The change of behavior occurs
when\cite{suzumura_confinement_ladder,tsuchiizu_confinement_ladder}
\begin{equation}
t_\perp^{\rm eff} = \Delta_{\rm 1ch}
\end{equation}
where $t_\perp^{\rm eff}$ is the renormalized interchain hopping.
We thus have to determine here whether we have two different
insulating phases. This can be done by looking at the fields that
order in each case. In the ``confined'' phase we showed in
section~\ref{sec:confined} that both $\phi_s$ and $\phi_a$ acquire
mean values. In the ``deconfined'' phase, as was shown in
section~\ref{sec:deconfined}, $\theta_\sigma$ and $\phi_\rho$
orders. The physical observable corresponding to an out of phase
charge density wave takes the following form with the two
different set of boson operators:
\begin{eqnarray}\label{jztranslation}
 J^z_{2k_f}&=&\psi^\dagger_1\psi_1-\psi^\dagger_2\psi_2 \\
&=&\frac{-2}{\pi a}\sin{\sqrt{2}\phi_s} \sin{\sqrt{2}\phi_a}\nonumber \\
&=&\frac{-2}{\pi a} \sin{\sqrt{2}\phi_\rho}\sin{\sqrt{2}\theta_\sigma} \nonumber
\end{eqnarray}
It is easy to see from (\ref{jztranslation}) that the
conditions $\langle \cos{\sqrt{8}\theta_\sigma}\rangle=-1$,
$\langle\cos{\sqrt{8} \phi_\rho}\rangle=-1$ on the one hand and
$\langle\cos{\sqrt{8}\phi_a}\rangle=-1$,
$\langle\cos{\sqrt{8} \phi_s}\rangle=-1$ on the other both give a non zero value to
$\langle J^z_{2k_f}\rangle$. Thus the two ``phases'' have the same
ordered fields (written in a different basis) and there is no
transition but a simple crossover. The out of phase charge density
waves ground state that arises naturally from the $K<1/2$ picture
and is shown in Figure~\ref{fig:cdwlock} stays valid even in the
other limit.

\subsection{Doped ladder}

Let us now determine the phase diagram of the ladder system away from half filling.
Introducing a chemical potential would change the Hamiltonian into
\begin{equation}
H = H_{\rm 1/2 filling} -\sqrt2\mu/\pi \int dx \nabla \phi_\rho
\end{equation}
As we have seen that the operator opening the charge gap is always
\begin{equation}
H_{\rm Mott ladder} = \int dx \cos(\sqrt8 \phi_\rho)
\end{equation}
instead of
\begin{equation}
H_{\rm Mott 1chain} = \int dx \cos(4 \phi_\rho)
\end{equation}
for the single chain. Using the known generic solution of such
Mott systems \cite{giamarchi_mott_ref} we can see that in the ladder the doping will destroy
the Mott phase when $\mu = \Delta$. The properties at the
transition are the generic properties of the Mott
transition in one dimension when varying the doping (Mott-$\delta$),
namely: (i) a dynamical exponent
$z=2$; (ii) a divergent compressibility on the metallic side. The
luttinger liquid exponent is universal and is $K^* = 1/2$.
This values is quite different from the one of a single chain.
The differences are recalled in Table~\ref{ta:luttinger}.
\begin{table}
\caption{\label{ta:luttinger} Luttinger liquid parameters at the Mott
transition, when varying the strength of the interaction (Mott-U)
and when varying the doping (Mott-$\delta$).}
\begin{ruledtabular}
\begin{tabular}{lll}
Parameter & 1chain & Ladder \\
$K^*_{\rm Mott-U}$      & 1/2 & 1 \\
$K^*_{\rm Mott-\delta}$ & 1/4 & 1/2
\end{tabular}
\end{ruledtabular}
\end{table}
The difference in $K^*$ should be observable in numerical calculations
such as exact diagonalization and dmrg. It might be a more clear signature
of the difference between the ladder and the single chain that the gap itself
specially for $K<1/2$.

These different values of $K^*$ correspond to two different kinds
of elementary excitations. For the gapped single spinless chain
the elementary excitations are domain walls separating charge
density waves with a different phase, the equivalent of the
excitations called spinons in the spin$-1/2$ Heisenberg model. In
a path-integral picture these corresponds to configurations where
the field $\phi$ is step-like with for instance
$\phi(-\infty,t)=0$ and $\phi(+\infty,t)=\pi/2$. However such a
configuration on one chain with a nearly constant configuration on
the second chain has an infinite cost when there is a
$\cos{\sqrt8\phi_\rho}$ term in the model, no matter how small the
coupling constant. Therefore the low-energy excitations allowed
for two coupled chains are kinks (two domain walls confined in the
same region). This behavior is reminiscent of the
double-sine-Gordon model that appears for a spontaneously
dimerized frustrated spin$-1/2$ chain (next-nearest-neighbor
exchange) with an infinitesimal bond alternation perturbation
 \cite{haldane_xxzchain}.

\begin{figure}
\centerline{\epsfig{file=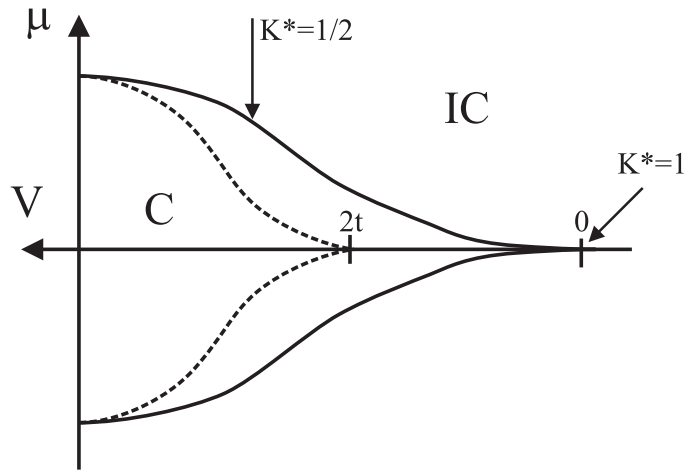,angle=0,width=\figwidth}}
\caption{\label{fig:summary} Phase diagram as a function of the
chemical potential $\mu$. The dashed line shows the
metal-insulator (commensurate-incommensurate) transition for a
single chain (beginning at finite interaction $V=2t$), the
continuous line corresponds to the spinless fermionic ladder.
$K^*$ is the universal value of the luttinger liquid parameter at
the transition. }
\end{figure}

A summary of the phase diagram is shown on
Figure~\ref{fig:summary}.

\section{Conclusion}\label{sec:conclusion}

We have investigated in this paper commensurate ladders of
spinless fermions. We have shown that contrarily to the single
chain, that needs a finite repulsion to become a Mott insulator, a
two leg ladder turns insulating for arbitrary small repulsion
between the fermions. The interchain hopping thus paradoxically
reinforces the insulating behavior of the system. We have computed
the charge gap as a function of the Luttinger exponent $K$ of a
single chain. For values of $K$ for which the single chain system
would be metallic ($1/2<K<1$) the charge gap vanishes as a power
law of the interchain hopping $t_\perp$ for small $t_\perp$, and
faster than the standard BKT behavior, usually characterizing
metal insulator transitions in one dimension, when the
interactions tend to zero. The interchain hopping also affects the
universal properties of the transition that occur upon doping the
system. Indeed for very small doping the Luttinger liquid
parameter takes the universal value $K^*=1/2$ instead of $K^*=1/4$
for a single chain (when a Mott gap is present). This difference
in Luttinger liquid parameters should be accessible in numerical
simulations such as exact-diagonalization or dmrg.

Given the fact that when a Mott gap is well established in a
single chain the interchain hopping scales to zero, whereas it
scales to large values otherwise, it was important to know whether
there was a confinement-deconfinement transition. We have shown
that there is in the spinless ladder only a crossover and that the
two seemingly different insulating phases (large Mott gap on a
single chain and small $t_\perp$ and large $t_\perp$ and small (or
zero) single chain Mott gap) are in fact of the same nature.

This leaves open how this cross-over should evolve for
an increasing number of coupled chains to give back the
confinement-deconfinement transition expected for an infinite number of chains.

\begin{acknowledgments}

T.G. and P.D. would like to thank D. Poilblanc and S. Capponi for
interesting discussions in the early stages of this work. One of
the author (Y.S.) is indebted to the financial support from
Universit{\'e} Paris--Sud. This work was partially supported by a
Grant-in-Aid for Scientific Research from the Ministry of
Education, Science, Sports and Culture (Grant No.09640429), Japan.
\end{acknowledgments}

\appendix
\section{terms generated to second order by $t_\perp$}\label{ap:alternative}
The RG equations (\ref{eq:flow2chain}) show that during the first
part of the flow up to $l_1$, $t_\perp$ generates to second order
the term $g_a\cos{\sqrt8\phi_a}$. Contracting $g_a$ with $g_u$
gives a contribution to the renormalization of the term
$g_s\cos{\sqrt8\phi_s}$ (noticing that $\phi_\rho=\phi_s$, we
switch to the notation $g_\rho$ instead of $g_s$). This term is
initially not present in the hamiltonian. However it is generated
and relevant for any repulsive interaction. One must thus make
sure that this term does not grow so fast as to open the gap
before the original umklapp term. If it were the case this would
change the expression for the charge gap compared to the one given
in the text.

To determine the growth of this coupling, we  have to determine
$g_a(l)$. Naively the scaling dimension of $g_a$ is $2-2K$, however it is
generated by $t_\perp^2$, term of dimension $4-(K+1/K)$. Comparing
the two rates shows that the source term dominates. Integration then shows that
$g_a(l)\propto t_\perp^2(l)$. We may now repeat the same analysis
for $g_\rho$, indeed its scaling dimension is $2-2K$ but it is
actually driven by $g_u(l)g_a(l)$, term of dimension $6-5K-1/K$,
dominant in the regime considered here ($K>1/2$). Consequently one
has
$g_\rho(l)\propto g_u(l)g_a(l)$. Thus at the scale $l_1$,
where $t_\perp$ is of order one, one has $g_\rho(l_1)\approx
g_u(l_1)$.

To complete the argument we note that the couplings
$g_\rho(l)\cos{\sqrt8\phi_\rho}$ and
$g_u(l)\cos{\sqrt8\phi_\rho}\cos{\sqrt8\theta_\sigma}$ are above
$l_1$ basically the same operator. Indeed in the $\sigma-$sector
$g_\sigma$ has flowed to strong coupling above $l_\sigma$,
$\langle\cos{\sqrt8\theta_\sigma}\rangle\neq 0$. Recalling that
the flow between $l_1$ and $l_\sigma$ does not change the
asymptotic dependence on $t_\perp$ and that $g_u(l_1)$ and
$g_\rho(l_1)$ are proportional we recover by following the flow of
the coupling $g_\rho$ the behavior explained in the text.

%\bibliography{totphys,patphys}

\end{document}